\documentclass[]{aa}

\def\eps@scaling{.95}
\def\epsscale#1{\gdef\eps@scaling{#1}}

\def\plotone#1{\centering \leavevmode
    \epsfxsize=\eps@scaling\columnwidth \epsfbox{#1}}

\def\plotfiddle#1#2#3#4#5#6#7{\centering \leavevmode
    \vbox to#2{\rule{0pt}{#2}}
    \includegraphics{#1}}

\usepackage{rotating}
\usepackage{graphicx}
\usepackage{latexsym}
\begin{document}
\thesaurus{11.01.1; 11.03.2; 11.04.1; 11.04.2; 11.05.2; 11.19.7; 12.12.1}
\title{A spectrophotometric catalogue of HII galaxies
\thanks{Based on observations obtained at the German-Spanish Observatory at
Calar Alto, Almeria, Spain}}
\author{Cristina C. Popescu \inst{1,2}
\and  Ulrich Hopp  \inst{3}
}

\offprints{Cristina C. Popescu (email address: popescu@levi.mpi-hd.mpg.de)}
\institute{Max Planck Institut f\"ur Kernphysik, Saupfercheckweg 1, 
           D--69117 Heidelberg, Germany
\and The Astronomical Institute of the Romanian Academy, Str. Cu\c titul de
Argint 5, 75212, Bucharest, Romania
\and Universit\"atssternwarte M\"unchen, Scheiner Str.1, 
       D--81679 M\"unchen, Germany}
\date{Received; accepted}
\maketitle
\markboth{C.C.Popescu \& U.Hopp}{A spectrophotometric catalogue of HII galaxies}
\begin{abstract}
We present a spectrophotometric catalogue of 90 emission-line galaxies (ELGs)
discovered during an objective - prism survey that aimed to search for dwarf
galaxies within the voids (Popescu et al. 1996, 1998). The paper presents line
ratios, equivalent widths and absolute fluxes for the emission-lines seen in 
the spectra of the galaxies. A list of newly discovered Wolf-Rayet galaxies is 
presented. Many objects included in the catalogue have low metallicity and 
the extreme cases are  proposed as candidates for very low metallicity 
galaxies. 
\end{abstract}


\section{Introduction}

Popescu et al. (1996, 1998) conducted a survey for emission-line galaxies 
(ELGs) based on the Hamburg Quasar Survey (HQS, Hagen et al. 1995) - 
IIIa-J digitised objective prism plates. The main goal of the project was to 
search for dwarf ELGs in voids and to analyse the large-scale structure of 
the distribution of this kind of galaxies (Popescu et al. 1997). The follow-up spectroscopy at 
the 2.2\,m telescope at the German-Spanish Observatory at Calar Alto 
(Almeria, Spain) provided us with a complete sample of 250 ELGs, of which 
many are Blue Compact Dwarfs (BCDs) or HII galaxies. These two terms are
usually used to define the same kind of objects. While the name BCD was mainly
used for objects classified on morphological criteria (Binggeli et al. 1985 -
for the Virgo Cluster Catalogue), the term of HII galaxy was introduced for
objects discovered on spectroscopic surveys for emission-line galaxies. 
These objects have focused attention after the discovery by Sargent \& 
Searle (1970) that some of them were low metallicity systems 
hosting a very active stellar formation. Gallagher \& Hunter (1989) showed that
the BCDs present also the lowest mass surface density and rotation velocities
ever measured for objects supporting star formation activity. Thus it was
suggested (Vilchez 1995) that these galaxies can be easily affected by
environmental factors. Since in our study of the spatial distribution of 
ELGs (Popescu et al. 1997) we identified the isolation of each sample galaxy,
we can address the question of environmental influences on the star-forming
properties of HII galaxies, using un unbiased sample. In this paper we give 
the catalogue of the
spectroscopic parameters of a subsample of 90 ELGs and we describe the data 
sample. The influence of the environment on the mechanism that controls star 
formation is discussed in a separate paper (Popescu et al. 1999).

\section{The data}
From our sample of 250 ELGs we selected for analysis only objects from a
region North of the \lq\lq Slice of the Universe\rq\rq (de Lapparent et al. 
1986), which is dominated by
very well defined nearby voids. This subsample consists of 90 ELGs, most of
them being distributed in the sheets and filaments that surround the voids, and
with a few ELGs within some nearby voids or at the rim of the voids. No rich
cluster of galaxies was included in the survey.

Most of the galaxies from this sample were observed twice. The objects were
first observed in a snap-shot mode (Popescu et al. 1997), in 
order to check the selection criteria and to measure redshifts. In these 
campaigns we mostly did not have photometric conditions, therefore no 
reliable absolute fluxes could have been derived. Also a slit width of
2$^{\prime\prime}$ was used, which did not provide a total coverage of the galaxies.
 The observing campaign from May 1997 (see Table 1 for the details of the
observations) was
\begin{table}[htp]
\caption[]{The details of the spectroscopic observations from May 1996}
\begin{tabular}{ll}
                      & \\
\hline\hline
                 &\\
Detector         & Lor-80 \\
Pixel Size ($\mu$)  & 15 \\
Slit width (${\prime\prime}$) & 4 \\
P.A.                  & 90. \\
Pixel number          & 1024$\times$1024 \\
Grism               & 9    \\
Dispersion (\AA/pixel) & 5.6 \\
Resolution (\AA) & 17 \\
Spectral Range (\AA)  & 3600-9000 \\
& \\
\hline
\end{tabular}
\end{table}
dedicated to the spectrophotometry of our objects, but it also completed 
the observations of selected candidates. These latter observations were done
in good photometric conditions, with a slit width of 4$^{\prime\prime}$. For
the small projected sizes of our dwarf galaxies, such an aperture is large
enough to encompass most of their line emission. For the purpose of estimating
the fraction of the line emission included in the slit we list in all the
relevant tables of the paper the seeing-corrected angular sizes of the
star-forming HII regions (r$_{HII}$). The comparison between the angular sizes
of the emitting regions and the aperture sizes lead to the conclusion that 
the fluxes measured in the campaign from May 1996 represent accurate 
measurements of the total integrated fluxes of our galaxies. The large 
aperture used to 
obtain total fluxes reduces the resolution of our spectra, therefore any 
deblending of the emission lines was performed on the spectra taken in the 
previous campaigns. For the few galaxies observed only in the May 1997 
observing campaign, no deblending was performed on the lines. Also, for the 
galaxies observed only in non-photometric conditions, no fluxes were 
assigned, only relative line ratios.

\begin{table}[htb]
{\small
\begin{tabular}{ll}
\multicolumn{2}{c}{{\bf Table 2 }}\\
\multicolumn{2}{c}{}\\
\hline\hline
& \\
error & emission line flux or equivalent width\\
\hline
&\\
 5$\%$ & H$\beta$, [OIII]\,${\lambda}$4959, [OIII]\,${\lambda}$5007, 
H$\alpha$, \\
       & W([OIII]), W(H$\alpha$)\\
10$\%$ & [OII]\,${\lambda}$3727, H$\gamma$, HeI\,${\lambda}$5876, W(H$\beta$)\\
15$\%$ & H$\delta$, [SII]\,${\lambda}$6717,6731\\
20$\%$ & [OI]\,${\lambda}$6300, HeI\,${\lambda}$6678, HeII\,${\lambda}$4686, \\
       & W([SII])\\
25$\%$ & HeI\,${\lambda}$7065, [AIII]\,${\lambda}$7136 \\
30$\%$ & [OI]\,${\lambda}$6364 \\
& \\
\hline
\end{tabular}
}
\end{table}

The data were reduced using the MIDAS routines (as described by Popescu et 
al. 1996, 1998).
The photometric calibrations of the spectra taken in the May 97 observing run
were done using the standard stars: HZ 21, HZ 44 and BD +33$^{\circ}$2642 (Oke
1990) and the absolute fluxes are accurate to about 10${\%}$. For one 
observing night the accuracy of the photometry was as good as 5$\%$.
Fluxes and equivalent widths of the emission lines were measured
interactively using the integrate/line routine in MIDAS. The errors in the
measurements were derived from repeated observations of the lines, and the
errors of the fluxes and equivalent widths W for different lines are given in
Table 2.

For all the galaxies with a strong underlying continuum emission (relative to
the  H$\beta$ line emission, W(H$\beta) < 20$\,$\AA$) we corrected the 
H$\beta$ fluxes for underlying stellar absorption with an assumed constant
equivalent width of 2\,$\AA$ (McCall et al. 1985). The H$\beta$ 
fluxes were afterwards corrected for reddening due to dust in our own Galaxy
and in the galaxy being observed. We used the reddening
coefficient c(H$\beta$), derived from the observed H$\alpha$/H$\beta$ Balmer 
line ratios. From equation (7.6) of Osterbrock (1974), the intrinsic ratio
I(H${\alpha}$)/I(H${\beta}$) is related to the observed ratio via 
\begin{eqnarray}
\frac {{\rm F}({\rm H}{\alpha})}{{\rm F}({\rm H}{\beta})} = 
\frac {{\rm I}({\rm H}{\alpha})}{{\rm I}({\rm H}{\beta})}\times
10^{-{\rm c}({\rm H}{\beta})[{\rm f}({\rm H}{\alpha})-{\rm f}({\rm H}
{\beta})]}
\end{eqnarray}\\
For the extinction in our Galaxy, f(H${\alpha})-{\rm f}({\rm H}{\beta})$ is 
the standard Galactic reddening law (Whitford 1958), with the extinction 
values taken from Burstein \& Heiles (1984). For the external extinction 
f(H${\alpha})-{\rm f}({\rm H}{\beta})$ is the reddening law 
given by Howarth (1983), which should be closer to the average reddening law
for the dwarf galaxies in our sample than the standard galactic one.    
We assumed that the intrinsic Balmer-line ratios are equal to the
case B recombination values of Brocklehurst (1971) for an electron 
temperature of 10$^4$\,K and an electron density of 100\,cm$^{-3}$.

For all the objects considered in this paper we also  obtained B and R
frames. Most of the direct images were either obtained for the purpose
of aquisition of the following spectroscopy or as a snap-shot survey to obtain total
magnitudes. Dedicated deeper (600 sec B and 300 sec R) images were observed for
all isolated ELGs. The photometry of the whole sample is discussed in detailed
in Vennik et al. (1999). Based on their photometric parameters (absolute
magnitudes, diameters) and the morphological appearance on the CCD images, the
galaxies were classified into the morphological classes proposed by Salzer et
al. (1989a,b), namely Star Burst Nucleus Galaxies (SBN), Dwarf Amorphous 
Nuclear Starburst Galaxies 
(DANS), HII Hotspot Galaxies (HIIH), Dwarf HII Hotspot Galaxies (DHIIH),
Sargent-Searle Objects (SS), Magellanic Irregular (Im), and Interacting Pairs 
(IP). Nevertheless, an independent check was done using the line ratios
of the emission-lines and their location in the diagnostic diagrams, and there
is an overall good agreement between the morphological and spectroscopic
classification.

\begin{table*}[hp]
{\small
\begin{tabular}{lccrrrrrccl}
\multicolumn{11}{c}{{\bf Table 3 }}\\
\multicolumn{11}{c}{}\\
\hline\hline
& & & & & & & & & & \\
(1) & (2) & (3) & (4) & (5) & (6) & (7) & (8) & (9) & (10) & (11)\\
Galaxy & M$_B$ & z & c(H$\beta$) & H$\beta$ & [OIII] & H$\alpha$ &
 [SII] &  F(H$\beta)_{obs}$ & r$_{HII}$ & type\\
& & & &W[\AA] &W[\AA] &W[\AA] &W[\AA] 
&[${\rm erg}\,{\rm s}^{-1}\,{\rm cm}^{-2}$] & [$^{\prime\prime}$] &\\
\hline
& & &  & & & & & & &\\
	    
HS1222+3741 & -17.67 & 0.0409 & 0.151 & 114 & 681 & 593 & 39 &1.67e-14&	0.6 & Im/BCD     \\	 
HS1223+3938 & -18.73 & 0.0360 & 0.474 &  15 &  91 & 101 & 14 &6.88e-15&	1.7 & HIIH       \\	 
HS1232+3846 & -19.70 & 0.0528 & 0.254 &  11 &  14 &  75 & 23 &4.18e-15&	2.0 & SBN        \\	 
HS1232+3947 & -17.16 & 0.0210 & 0.286 &  29 & 139 & 172 & 13 &8.90e-15&	1.7 & DHIIH      \\	 
HS1232+3612 & -19.77 & 0.0425 & 0.411 &  18 &  85 & 136 & 25 &1.27e-14&	    & IP         \\	 
HS1236+3821 & -17.06 & 0.0073 & 0.602 &   4 &  23 &  37 &  9 &7.72e-15&	2.1 & DHIIH      \\	 
HS1236+3937 & -15.67 & 0.0184 & 0.081 & 112 & 506 & 521 & 56 &5.72e-15&	0.8 & DHIIH/SS   \\	 
HS1240+3755 & -21.1  & 0.0860 & 0.969 &   3 &   6 &  64 & 14 &	 -    &	    & IP/SBNpec        \\	 
HS1244+3648 & -18.82 & 0.0472 & 0.000 &  35 & 146 & 140 & 38 &1.52e-14&	0.8 & HIIH       \\	 
HS1256+3505 & -18.75 & 0.0342 & 0.448 &  20 &  55 & 155 & 32 &1.60e-14&	1.1 & DANS       \\	 
HS1258+3438 & -16.4  & 0.0248 & 0.380 &  36 & 209 & 189 & 15 &4.39e-15&	    & DHIIH      \\	 
HS1301+3312 & -17.41 & 0.0371 & 0.391 &  31 & 125 & 177 & 40 &5.90e-15&	0.8 & IP         \\	 
HS1301+3325 & -16.96 & 0.0246 & 0.573 &  11 &  51 &  73 &  8 &3.71e-15&	1.3 & DHIIH      \\	 
HS1301+3209 & -17.06 & 0.0238 & 0.579 &  10 &  36 &  78 & 13 &3.01e-15&	2.2 & HIIH       \\	 
HS1304+3529 & -17.38 & 0.0165 & 0.013 & 118 & 570 & 617 & 76 &4.82e-14&	1.9 & IP         \\	 
HS1306+3320 & -18.32 & 0.0270 & 0.508 &  36 & 116 & 199 & 27 &1.32e-14&	1.9 & HIIH       \\	 
HS1308+3044 & -17.92 & 0.0209 & 0.564 &   4 &  13 &  40 &  8 &4.55e-15&	2.5 & DANS       \\	 
HS1311+3628 & -16.84 & 0.0031 & 0.160 & 301 &1550 &1834 &139 &	    - &	    & DHIIH      \\	 
HS1312+3847 & -18.77 & 0.0515 & 0.037 &  74 & 337 & 392 & 58 &1.53e-14&	1.4 & HIIH       \\	 
HS1312+3508 & -16.26 & 0.0035 & 0.164 & 254 &1466 &1463 &179 &1.02e-13&	    & DHIIH      \\	 
HS1315+3132 & -16.46 & 0.0315 & 0.476 &  36 & 136 & 175 & 28 &4.14e-15&	0.8 & DHIIH      \\	 
HS1318+3239 & -17.22 & 0.0435 & 0.186 &  79 & 464 & 394 & 32 &5.75e-15&	1.2 & IP         \\	 
HS1319+3224 & -15.3  & 0.0182 & 0.208 &  49 & 241 & 237 & 12 &	   -  &	0.8 & SS/DHIIH         \\	 
HS1325+3225 & -17.9  & 0.0504 & 0.542 &  39 & 102 & 208 & 31 &	   -  &	    & DHIIH/HIIH \\	 
HS1325+3255 & -15.92 & 0.0263 & 0.146 &  74 & 412 & 340 & 14 &4.38e-15&	0.8 & DHIIH/SS   \\	 
HS1327+3126 & -18.07 & 0.0568 & 0.192 &  91 & 514 & 484 & 42 &1.07e-14&	0.6 & DHIIH      \\	 
HS1328+3424 & -17.79 & 0.0227 & 0.503 &   8 &  18 &  54 &  4 &2.15e-15&	    & HIIH       \\	 
HS1329+3703 & -19.11 & 0.0557 & 0.214 &   7 &   9 &  54 & 15 &	   -  &	1.9 & HIIH/DANS  \\	 
HS1330+3651 & -17.15 & 0.0167 & 0.147 &  72 & 387 & 371 & 45 &2.62e-14&	    & DHIIH      \\	 
HS1332+3426 & -16.14 & 0.0220 & 0.444 &  35 & 167 & 195 & 17 &4.27e-15&	1.4 & DHIIH      \\	 
HS1334+3957 & -15.32 & 0.0083 & 0.000 &  71 & 351 & 334 & 26 &	   -  &	1.8 & DHIIH      \\	 
HS1336+3114 & -17.94 & 0.0158 & 0.000 &   5 &  11 &  29 & 14 &	   -  &	3.9 & HIIH       \\	 
HS1340+3307 & -16.63 & 0.0158 & 0.539 &  20 & 108 & 128 & 23 &9.73e-15&	1.6 & DHIIH      \\	 
HS1341+3409 & -16.35 & 0.0171 & 0.480 &  15 &  59 &  85 & 19 &5.34e-15&	1.2 & DHIIH      \\	 
HS1347+3811 & -15.30 & 0.0103 & 0.243 &  64 & 363 & 290 & 33 &8.62e-15&	4.9 & DHIIH      \\	 
HS1349+3942 & -15.16 & 0.0054 & 0.337 &  13 &  45 &  85 & 23 &1.05e-14&	1.8 & DHIIH      \\	 
HS1354+3634 & -17.10 & 0.0167 & 0.513 &  19 &  50 & 123 & 26 &1.12e-14&	1.2 & DANS       \\	 
HS1354+3635 & -17.81 & 0.0171 & 0.507 &  21 &  76 & 124 & 25 &2.04e-14&	3.3 & HIIH       \\	 
HS1402+3650 & -18.81 & 0.0347 & 0.561 &  26 &  85 & 192 & 43 &1.36e-14&	1.6 & HIIH       \\	 
HS1410+3627 & -18.11 & 0.0338 & 0.311 &  15 &  55 &  97 & 28 &5.16e-15&	    & HIIH       \\	 
HS1413+4402 & -19.67 & 0.0698 & 0.607 &  10 &   5 &  99 & 13 &	   -  &	1.3 & SBN        \\	 
HS1416+3554 & -16.94 & 0.0103 & 0.328 &  14 &  49 &  91 & 24 &7.70e-15&	3.4 & HIIH/DHIIH \\	 
HS1420+3437 & -16.75 & 0.0246 & 0.448 &  27 &  56 & 149 & 17 &7.53e-15&	0.6 & DHIIH      \\	 
HS1422+3325 & -17.30 & 0.0341 & 0.337 &  20 &  50 & 119 & 30 &4.69e-15&	0.8 & HIIH       \\	 
HS1422+3339 & -16.4  & 0.0114 & 0.350 &  17 &  58 &  95 & 20 &1.12e-14&	1.8 & DHIIH      \\	 
HS1424+3836 & -16.06 & 0.0218 & 0.195 & 104 & 579 & 545 & 24 &1.00e-14&	1.4 & DHIIH?     \\	 
HS1425+3835 & -17.80 & 0.0223 & 0.099 &  12 &  24 &  64 &  9 &	    - &	1.8 & HIIH       \\	 
HS1429+4511 & -17.40 & 0.0321 & 0.307 &  15 &   2 &  96 & 32 &	    - &	1.0 & DANS       \\	 
HS1429+3154 & -16.73 & 0.0117 & 0.143 &  34 & 144 & 162 & 26 &2.06e-14&	2.6 & DHIIH      \\	 
HS1438+3147 & -17.88 & 0.0443 & 0.308 &  35 & 188 & 198 & 30 &6.54e-15&	1.2 & DANS       \\	 
HS1440+4302 & -15.07 & 0.0085 & 0.336 &  44 & 269 & 238 & 31 &1.13e-14&	1.2 & DHIIH/SS   \\	 
HS1440+3120 & -17.4  & 0.0525 & 0.221 & 152 & 925 & 845 & 62 &9.60e-15& 0.7 & DHIIH   \\	 
HS1440+3805 & -18.80 & 0.0322 & 0.263 &   7 &  14 &  46 & 17 &	    - &	2.9 & HIIH       \\	 
& & & & & & & & & &\\
\hline
\end{tabular}
}
\end{table*}
\begin{table*}[hp]
{\small
\begin{tabular}{lccrrrrrccl}
\multicolumn{11}{c}{{\bf Table 3 (continued)}}\\
\multicolumn{11}{c}{}\\
\hline\hline
& & & & & & & & & &\\
(1) & (2) & (3) & (4) & (5) & (6) & (7) & (8) & (9) & (10) & (11)\\
Galaxy & M$_B$ & z & c(H$\beta$) & H$\beta$ & [OIII] & H$\alpha$ &
 [SII] &  F(H$\beta)_{obs}$ & r$_{HII}$ & type\\
& & & &W[\AA] &W[\AA] &W[\AA] &W[\AA] 
&[${\rm erg}\,{\rm s}^{-1}\,{\rm cm}^{-2}$] & [$^{\prime\prime}$] &\\
\hline
& & &  & & & & & & &\\
HS1442+4250 & -14.73 & 0.0025 & 0.081 & 113 & 550 & 574 & 17 &2.95e-14&	    & SS         \\	 
HS1444+3114 & -19.02 & 0.0297 & 0.413 &  26 &  60 & 174 & 38 &2.78e-14&	1.7 & DANS       \\	 
HS1502+4152 & -16.66 & 0.0164 & 0.515 &   7 &  39 &  55 & 21 &2.22e-15&	    & DHIIH      \\	 
HS1507+3743 & -17.58 & 0.0322 & 0.114 & 232 &1374 &1307 & 68 &3.45e-14&	1.0 & DHIIH      \\	 
HS1529+4512 & -17.05 & 0.0231 & 0.222 &  16 &  99 & 102 & 12 &2.95e-15&	1.7 & HIIH/DHIIH \\	 
HS1544+4736 & -17.11 & 0.0195 & 0.025 &  32 & 154 & 162 & 30 &1.00e-14&	2.6 & DHIIH      \\	 
HS1546+4755 & -17.78 & 0.0377 & 0.485 &  35 & 181 & 204 & 27 &7.77e-15&	0.7 & DHIIH      \\	 
HS1609+4827 & -17.6  & 0.0096 & 0.381 &  12 &  40 &  86 & 20 &1.91e-14&	2.5 &  HIIH     \\	 
HS1610+4539 & -17.2  & 0.0196 & 0.480 &  26 & 106 & 152 & 24 &1.41e-14&	1.3 & DHIIH/HIIH \\	 
HS1614+4709 & -13.6  & 0.0026 & 0.100 & 140 & 880 & 915 & 55 &6.62e-14&	1.6 & SS      \\	 
HS1633+4703 & -15.93 & 0.0086 & 0.364 &  15 &  55 &  83 & 19 &1.34e-14&	1.1 & DHIIH      \\	 
HS1640+5136 & -19.59 & 0.0308 & 0.443 &  21 &  55 & 167 & 34 &2.42e-14&	2.0 & SBN/HIIH   \\	 
HS1641+5053 & -19.05 & 0.0292 & 0.574 &  15 &  65 & 110 & 21 &2.11e-14&	    & HIIH       \\	 
HS1645+5155 & -19.11 & 0.0286 & 0.330 &  47 & 191 & 265 & 45 &1.53e-14&	5.5 & HIIH       \\	 
HS1657+5735 & -20.60 & 0.0505 & 0.455 &  43 & 167 & 253 & 46 &5.43e-14&	1.2 & SBN        \\	 
HS1723+5631A& -18.70 & 0.0286 & 0.341 &  23 & 100 & 121 & 20 &1.13e-14&	    & IP         \\	 
HS1723+5631B& -18.70 & 0.0286 & 0.337 &  32 & 140 & 188 & 32 &1.16e-14&	    & IP         \\	 
HS1728+5655 & -16.38 & 0.0160 & 0.094 & 106 & 514 & 522 & 36 &2.39e-14&	0.9 & DHIIH      \\	 
& & & & & & & & & &\\
\hline
\end{tabular}
}
\end{table*}

\begin{table*}[hp]
\begin{sideways}
{\small
\begin{tabular}{lrrrrrrrrrrrrrrrrr}
\multicolumn{18}{c}{{\bf Table 4 }}\\
\multicolumn{18}{c}{}\\
\hline\hline
& & & & & & & & & &&&&&&&&\\
(1) & (2) & (3) & (4) & (5) & (6) & (7) & (8) & (9) & (10) & (11) &
(12) & (13) & (14) & (15) & (16) & (17) & (18) \\
Galaxy & [OII] & [NeIII] &H$\delta$&H$\gamma$&
 [OIII]&   [OIII] & [OIII] &  HeI &  [OI] & [OI] & [NII] & HeI & 
[SII] & HeI &  [AIII] & [OII] & [AIII]\\
& 3727 & 3869 & 4102 & 4340 & 4363 & 4959 & 5007 & 5876 & 6300 & 6364 &
6584 & 6678 & 6717 & 7065 & 7136 & 7320 & 7751\\
& & & & & & & & & &&&&6731&&&7330&\\
\hline
& & &  & & & & & & &&&&&&&&\\
HS1222+3741 & 168 & 68  & 22 & 45 & 15  &  185&  556 & 10 &  4 & - &   - & - & 16 & - &  5&  -  &  -\\
HS1223+3938 & 379 & 75  &  - & 36 &  -  &  177&  509 & 17 &  - & - &   - & - & 38 & - &  3&  -  &  -\\
HS1232+3846 & 334 &  -  &  - & 32 &  -  &   38&  103 &  - &  - & - &  68 & - & 109 & - &-  &  -  &  -\\
HS1232+3947 & 256 &  -  &  - & 34 &  -  &  140&  427 & 14 & 16 & - &   5 & 9 & 21 & - &  -&  -  &  -\\
HS1232+3612 & 428 & 69  &  - & 43 &  -  &  132&  399 &  9 & 13 & - &  27 & 7 &  56 & - & 4 &  -  &  -\\
HS1236+3821 & 849 &  -  &  - &  - &  -  &  127&  380 & 11 &  8 & - &  19 & - & 73 & - &  -&  -  &  -\\
HS1236+3937 & 116 & 78  & 27 & 51 & 14  &  152&  445 & 13 &  - & - &   - & - & 26 & - &  -&  -  &  -\\
HS1240+3755 & 659 &  -  &  - &  - &  -  &   76&   94 &  - &  - & - & 179 & - &  96 & - & - &  -  &  -\\
HS1244+3648 & 292 & 55  &  - & 41 &  -  &  135&  406 & 15 & 16 & 7 &   - & - & 72 & - &  5&  -  &  -\\
HS1256+3505 & 467 &  -  &  - & 60 &  -  &   87&  251 & 15 &  7 & - &  63 & - & 73 & - &  9&  -  &  -\\
HS1258+3438 & 222 & 43  & 23 & 48 &  2  &  193&  534 & 10 &  - & - &   - & - & 22 & - &  -&  -  &  -\\
HS1301+3312 & 439 &  -  &  - & 44 &  -  &  131&  393 &  9 & 17 & - &   - & - &  58 & - &-  &  -  &  -\\
HS1301+3325 & 508 &  -  &  - &  - &  -  &  125&  350 &  - &  - & - &   - & - &  28 & - & - &  -  &  -\\
HS1301+3209 & 912 &  -  &  - &  - &  -  &   99&  287 &  - & 19 & - &  10 & - &  52 & - &-  &  -  &  -\\
HS1304+3529 & 185 & 55  & 24 & 46 & 10  &  152&  451 & 10 &  9 & - &  10 & 1 & 33 & - &  6&  -  &  -\\
HS1306+3320 & 354 & 53  &  - & 50 &  -  &  108&  320 & 11 &  7 & - &   - & - & 37 & - &  8&  -  &  -\\
HS1308+3044 & 577 &  -  &  - &  - &  -  &   70&  198 &  - &  - & - &  59 & - &  68 & - &-  &  -  &  -\\
HS1311+3628 & 216 & 29  & 30 & 54 &  -  &  176&  492 & 11 &  4 & - &  13 & 3 & 28 & - &  9&  3  &  -\\
HS1312+3847 & 242 & 64  & 18 & 46 &  5  &  153&  445 & 18 &  6 & - &  10 & - & 47 & - &  7&  -  &  -\\
HS1312+3508 & 261 & 64  & 29 & 53 &  7  &  170&  492 & 12 &  5 & 3 &  11 & 5 & 30 & 4 & 11&  5  &  4\\
HS1315+3132 & 410 & 20  & 24 & 50 &  5  &  123&  356 & 11 & 10 & - &   - & - & 42 & - &  8&  -  &  -\\
HS1318+3239 & 156 & 63  & 27 & 44 & 12  &  198&  577 & 13 &  - & - &   - & 3 & 23 & - &  6&  -  &  -\\
HS1319+3224 & 124 & 63  & 26 & 48 & 13  &  163&  503 & 10 &  - & - &   - & - & 14 & - &  -&  -  &  -\\
HS1325+3225 & 353 &  -  &  - & 44 &  -  &   84&  250 & 10 &  - & - &   - & 3 & 40 & - &  -&  -  &  -\\
HS1325+3255 & 126 & 57  & 22 & 49 &  8  &  187&  548 &  9 &  6 & - &   - & - &  12 & - & - &  -  &  -\\
HS1327+3126 & 199 & 56  & 26 & 46 &  7  &  182&  536 & 12 &  5 & 4 &   - & 3 & 25 & 3 &  5&  3  &  2\\
HS1328+3424 & 434 &  -  &  - &  - &  -  &   84&  183 &  - &  - & - &  34 & - &  25 & - & - &  -  &  -\\
HS1329+3703 & 366 &  -  &  - &  - &  -  &   35&  102 &  - &  - & - &  85 & - & 111 & - & - &  -  &  -\\
HS1330+3651 & 186 & 48  & 25 & 43 &  9  &  164&  490 & 15 &  8 & - &   9 & - & 35 & - &  4&  -  &  -\\
HS1332+3426 & 359 & 74  &  - & 53 &  -  &  162&  455 & 14 & 11 & - &   - & - & 23 & - &  6&  2  &  -\\
HS1334+3957 & 288 & 48  & 33 & 55 &  -  &  178&  519 & 17 &  - & - &   5 & - & 24 & - &  7&  -  &  -\\
& & & & & & & & & & & & & & & & &\\
\hline
\end{tabular}
}
\end{sideways}
\end{table*}
\begin{table*}[hp]
\begin{sideways}
{\small
\begin{tabular}{lrrrrrrrrrrrrrrrrr}
\multicolumn{18}{c}{{\bf Table 4 (continued)}}\\
\multicolumn{18}{c}{}\\
\hline\hline
& & & & & & & & & & &&&&&&&\\
(1) & (2) & (3) & (4) & (5) & (6) & (7) & (8) & (9) & (10) & (11) &
(12) & (13) & (14) & (15) & (16) & (17) & (18) \\
Galaxy & [OII] & [NeIII] &H$\delta$&H$\gamma$&
 [OIII]&  [OIII] & [OIII] &  HeI &  [OI] & [OI] & [NII] & HeI & 
[SII] & HeI &  [AIII] & [OII] & [AIII]\\
& 3727 & 3869 & 4102 & 4340 & 4363 & 4959 & 5007 & 5876 & 6300 & 6364 &
6584 & 6678 & 6724 & 7065 & 7136 & 7320 & 7751\\
& & & & & & & & & &&&&6731&&&7330&\\
\hline
& & &  & & & & & & &&&&&&&&\\
HS1336+3114 & 278 &  -  &  - &  - &  -  &   79&  172 &  - &  - & - &  51 & - & 163 & - & - &  -  &  -\\
HS1340+3307 & 503 &153  &  - & 66 &  -  &  174&  485 & 16 &  - & - &   7 & - & 49 & - &  -&  -  &  -\\
HS1341+3409 & 573 & 88  &  - &  - &  -  &  126&  325 & 18 &  - & - &   - & - & 60 & - &  -&  -  &  -\\
HS1347+3811 & 229 & 90  & 28 & 54 &  7  &  189&  546 & 13 &  8 & - &   - & - & 33 & 3 &  8&  -  &  -\\
HS1349+3942 & 556 &  -  &  - & 53 &  -  &  106&  286 & 13 & 12 & - &  26 & - & 82 & - &  -&  -  &  -\\
HS1354+3634 & 467 &  -  &  - & 48 &  -  &   91&  255 & 10 & 18 & - &  44 & - & 71 & - &  -&  -  &  -\\
HS1354+3635 & 468 &  -  &  - & 48 &  -  &  118&  339 & 10 &  7 & - &  28 & 3 & 61 & - &  4&  -  &  -\\
HS1402+3650 & 483 & 85  &  - & 49 &  -  &   96&  294 & 12 & 13 & - &  20 & 4 & 63 & - &  -&  -  &  -\\
HS1410+3627 & 601 &  -  &  - & 46 &  -  &  109&  305 & 14 & 14 & - &  14 & - & 81 & - &  -&  -  &  -\\
HS1413+4402 & 214 &  -  &  - &  - &  -  &   26&   45 &  - &  - & - & 150 & - &  62 & - &-  &  -  &  -\\
HS1416+3554 & 593 & 87  &  - &  - &  -  &   99&  284 &  - &  - & - &  18 & - &  77 & - &-  &  -  &  -\\
HS1420+3437 & 484 &  -  &  - & 49 &  -  &   79&  196 & 11 &  - & - &  22 & - & 34 & - &  -&  -  &  -\\
HS1422+3325 & 479 &  -  &  - & 50 &  -  &   90&  245 &  9 &  - & - &  59 & - &  86 & - & - &  -  &  -\\
HS1422+3339 & 466 &  -  &  - & 40 &  -  &  105&  289 &  - &  - & - &  21 & - &  65 & - &-  &  -  &  -\\
HS1424+3836 & 115 & 36  & 26 & 53 &  7  &  181&  552 & 12 &  4 & - &   - & - & 12 & - &  3&  -  &  -\\
HS1425+3835 & 586 &  -  &  - &  - &  -  &   59&  200 &  - &  - & - &  40 & - &  51 & - & - &  -  &  -\\
HS1429+4511 & 205 &  -  &  - &  - &  -  &   11&   51 &  - &  - & - &  82 & - & 112 & - & - &  -  &  0\\
HS1429+3154 & 392 &  -  &  - & 46 &  -  &  142&  419 &  6 &  6 & - &  19 & - &  49 & - & 7 &  -  &  -\\
HS1438+3147 & 335 & 71  &  - & 43 &  -  &  174&  507 & 14 &  8 & - &   7 & - & 42 & - &  -&  -  &  -\\
HS1440+4302 & 530 & 94  & 26 & 60 & 14  &  191&  558 & 10 &  6 & - &  11 & - & 36 & - &  7&  -  &  -\\
HS1440+3120 & 156 & 59  & 30 & 44 & 11  &  211&  613 & 13 &  5 & - &   - & 3 & 20 & - &  4&  -  &  -\\
HS1440+3805 & 397 &  -  &  - &  - &  -  &   78&  159 & 13 & 23 & - &  50 & - & 118 & - & -&  -  &  -\\
HS1442+4250 &  97 & 50  & 22 & 38 & 11  &  165&  475 & 12 &  - & - &   - & 2 &  8 & - &  4&  -  &  -\\
HS1444+3114 & 492 & 30  &  - & 50 &  -  &   75&  220 &  9 &  9 & 5 &  43 & - &  71 & - & 6 &  -  &  -\\
HS1502+4152 &   - &  -  &  - &  - &  -  &  153&  420 &  - &  - & - &  22 & - & 108 & - & - &  -  &  -\\
HS1507+3743 &  87 & 66  & 29 & 52 & 12  &  217&  658 & 10 &  3 & - &   - & 2 & 13 & 4 &  4&  4  &  -\\
HS1529+4512 & 465 & 76  &  - & 46 &  -  &  173&  502 &  - &  - & - &  16 & - &  34 & - & - &  -  &  0\\
HS1544+4736 & 104 &  -  &  - & 62 &  -  &  147&  456 & 15 &  7 & - &  14 & - & 52 & - &  7&  -  &  -\\
HS1546+4755 & 397 & 73  & 40 & 52 &  -  &  166&  494 & 10 &  7 & - &   6 & - & 37 & - &  -&  -  &  -\\
& & & & & & & & & & & & & & & & &\\
\hline
\end{tabular}
}
\end{sideways}
\end{table*}
\begin{table*}[hp]
\begin{sideways}
{\small
\begin{tabular}{lrrrrrrrrrrrrrrrrr}
\multicolumn{18}{c}{{\bf Table 4 (continued) }}\\
\multicolumn{18}{c}{}\\
\hline\hline
& & & & & & &  & & &&&&&&&&\\
(1) & (2) & (3) & (4) & (5) & (6) & (7) & (8) & (9) & (10) & (11) &
(12) & (13) & (14) & (15) & (16) & (17) & (18) \\
Galaxy & [OII] & [NeIII] &H$\delta$&H$\gamma$&
 [OIII]&  [OIII] & [OIII] &  HeI &  [OI] & [OI] & [NII] & HeI & 
[SII] & HeI &  [AIII] & [OII] & [AIII]\\
& 3727 & 3869 & 4102 & 4340 & 4363 & 4959 & 5007 & 5876 & 6300 & 6364 &
6584 & 6678 & 6724 & 7065 & 7136 & 7320 & 7751\\
& & & & & & & & &&&&&6731&&&7330&\\
\hline
& & &  & & & &  & &&&&&&&&&\\
HS1609+4827 & 551 &  -  &  - & 38 &  -  &   91&  261 &  8 & 15 & - &   - & - &  63 & - & - &  -  &  -\\
HS1610+4539 & 390 &  -  & 31 & 47 &  9  &  136&  374 & 10 & 10 & - &   - & 4 & 42 & - &  4&  -  &  -\\
HS1614+4709 &  79 & 54  & 27 & 50 &  6  &  200&  590 & 10 &  4 & - &   - & - & 18 & - &  5&  -  &  -\\
HS1633+4703 & 596 & 86  &  - & 48 &  -  &  116&  309 & 15 & 10 & - &  19 & 9 & 67 & - &  7&  -  &  -\\
HS1640+5136 & 520 & 22  &  - & 56 &  -  &   84&  243 & 15 & 11 & 4 &  98 & - & 79 & - &  5&  -  &  -\\
HS1641+5053 & 582 & 94  &  - & 33 &  -  &  121&  339 & 11 & 11 & - &   - & - & 50 & - &  -&  -  &  -\\
HS1645+5155 & 349 &  -  & 24 & 45 &  9  &  129&  387 & 22 &  7 & - &  12 & - & 46 & - &  -&  -  &  -\\
HS1657+5735 & 349 & 39  & 22 & 47 &  3  &  117&  355 & 10 &  7 & 1 &   - & 2 & 47 & - &  3&  -  &  2\\
HS1723+5631A& 425 &  -  &  - & 61 &  -  &  140&  412 & 15 & 14 & 6 &  23 & - & 49 & - &  -&  -  &  -\\
HS1723+5631B& 352 &  -  &  - & 54 &  -  &  144&  419 & 13 &  9 & - &  23 & - & 50 & - &  -&  -  &  -\\
HS1728+5655 & 211 & 75  & 26 & 47 &  3  &  172&  500 & 12 &  5 & 2 &  25 & 2 & 23 & - &  7&  3  &  -\\
& & & & & & & & & & & & & & & & & \\
\hline
\end{tabular}
}
\end{sideways}
\end{table*}

\begin{table}[]
\begin{sideways}
{\small
\begin{tabular}{lccrrrrcrrrrrrrrrrl}
\multicolumn{19}{c}{{\bf Table 5 }}\\
\multicolumn{19}{c}{}\\
\hline\hline
& & & & & & & & & & & & & & & & & \\
(1) & (2) & (3) & (4) & (5) & (6) & (7) & (8) & (9) & (10) & 
(11) & (12) & (13) & (14) & (15) & (16) & (17) & (18) & (19)\\
Galaxy & M$_B$ & z & H$\beta$ & [OIII] & H$\alpha$ & [SII] & F(H$\beta)_{obs}$ &
[OII]& H$\gamma$ & [OIII] & [OIII] & HeI & [OI] & H${\alpha}^{*}$ & HeI 
& [SII] & r$_{HII}$& type\\
 & & &W  & W &W &W &  & 3727 
& 4340 & 4959 & 5007 & 5876 & 6300 & 6563 & 6678 & 6724 &  & \\
& & &[\AA] &[\AA] &[\AA] &[\AA] &[${\rm erg}\,{\rm s}^{-1}\,{\rm cm}^{-2}$] & 
& & & &  & &6584 & &6731 &[$^{\prime\prime}$] &\\
\hline
& & & &  & & & & & & & & & & & & & &\\ 
HS1255+3506 &-16.61 & 0.0155 & 10 &  31 &  66 & 19 & 4.70e-15 & 381 &  - &  86 
& 258 &  - & 24 & 457 &  -  & 124 & 1.7 & DHIIH?\\  
HS1318+3406 &-18.65 & 0.0352 &  8 &  17 &  69 & 21 & 5.09e-15 & 414 &  - &  74 
& 178 &  - &  - & 526 &  -  & 152 &  & HIIH?\\   
HS1327+3412 & -     & 0.2510 & 11 &   7 & 111 & 11 &      -   &  87 &  - &  13 
&  52 &  - &  - & 376 &  -  &  37 & 0.7 &   -   \\  
HS1333+3149 &-18.61 & 0.0248 &  8 &  27 &  66 & 11 & 8.86e-15 & 387 &  - & 101 
& 267 &  - &  - & 470 &  -  &  77 & 2.3 & HIIH\\    
HS1340+3207 &-18.35 & 0.0365 & 18 & 105 & 127 & 23 & 6.92e-15 & 390 & 16 & 186 
& 555 & 21 &  - & 426 &  -  &  73 & 1.7 & DANS\\    
HS1429+3154$^{*}$ &-16.73 & 0.0117 & 34 & 144 & 162 & 26 & 2.06e-14 & 347 & 39 & 143 
& 424 &  6 &  7 & 344 &  -  &  55 & 2.6 & DHIIH\\   
HS1442+4332 &-20.05 & 0.0811 &  5 &   9 &  65 & 15 & 2.55e-15 & 361 &  - &  57 
& 116 & 15 & 43 & 616 & 14  & 126 & 1.5  & SBN?\\    
HS1543+4525 &-19.17 & 0.0389 & 10 &  15 &  92 & 27 & 6.90e-15 & 388 &  - &  38 
& 123 & 15 & 14 & 573 &  8  & 164 & 2.2 & DANS?\\   
HS1627+5239$^*$ &-17.88 & 0.0288 & 14 &  37 &  95 & 16 & 7.61e-15 & 286 &  - &  73 
& 224 & 12 &  - & 422 &  -  &  70 &1.22 & HIIH?\\   
HS1657+5033 &-17.43 & 0.0102 &  7 &  29 &  60 & 14 & 1.19e-14 & 338 &  - & 103 
& 323 & 10 &  - & 559 &  6  & 124 & 1.88 & HIIH\\    
& & & & & & & & & & & & & & & & & &\\
\hline
\multicolumn{19}{l}{}\\
\multicolumn{19}{l}{The galaxies HS1429+3154 and HS1627+5239 have detected 
[AIII]\,${\lambda}$7136, with the line ratios relative to H${\beta}$ of 8 and 
15, respectively.}\\
\end{tabular}
}
\end{sideways}
\end{table}
\begin{table}[]
\begin{sideways}
{\small
\begin{tabular}{lcccrrrccl}
\multicolumn{10}{c}{{\bf Table 6 }}\\
\multicolumn{10}{c}{}\\
\hline\hline
& & & & & & & & &\\
(1) & (2) & (3) & (4) & (5) & (6) & (7) & (8) & (9) & (10)\\
Galaxy    &  M$_B$ & z & F(H${\alpha})_{obs}$ & [OII] & [OIII] & [NII] 
& [SII] & r$_{HII}$ & type \\
& & & &3727 & 5007 & 6584 & 6724 & \\
& & & [${\rm erg}\,{\rm s}^{-1}\,{\rm cm}^{-2}$]     &      &      &  &6731 
& [$^{\prime\prime}$]\\
\hline
HS1309+3409 & -20.19 & 0.0785 & 1.59e-14 &   - &   -  &	41 &  14 & 1.5 & SBN \\
HS1331+3906 & -20.37 & 0.0643 & 1.11e-14 &   - &   -  &	37 &  33 & 2.4 & SBN \\
HS1336+3650 & -16.84 & 0.0202 & 3.16e-15 &   - & 114  &	 - &  -  & 1.5 & DHIIH\\ 
HS1421+4018 &  -     & 0.0982 & 6.12e-15 & 132 &   -  &	 - &  84 &   \\
HS1435+4523 & -20.29 & 0.1267 & 8.69e-15 &  60 &  22  &	26 &  65 & 1.9 & SBN\\
HS1505+3944 & -18.47 & 0.0366 & 5.20e-15 &  40 &   -  &	 - &  -  & 2.8 & DANS/IP\\
HS1522+4214 & -17.76 & 0.0190 & 3.74e-15 &  79 &  31  &	29 &  55 & 1.85 & DANS/DHIIH\\
& & & & & & & & &\\
\hline
\end{tabular}
}
\end{sideways}
\end{table}


\section{The catalogue}

The Catalogue is presented in Table 3 through Table 6. Table 3 and 4 present
the spectroscopic parameters of all the galaxies for which a reliable internal
extinction correction was done, and which account for  the main body of the 
Catalogue. 
The data are arranged in Table 3 as follows:
\begin{itemize}
\item
Col. 1: The galaxy name according to Popescu et al. (1996,1998).
\item
Col. 2: The absolute blue magnitudes (M$_B$) as listed in Vennik et al. (1999).
\item
Col. 3: The heliocentric redshift.
\item
Col. 4: The absorption coefficient c(H${\beta}$).
\item
Col. 5, 6, 7, 8: The equivalent widths (W) of the H${\beta}$, 
[OIII]\,${\lambda}$5007, H${\alpha}$ and [SII]\,${\lambda}{\lambda}$6717,6731 
emission lines in $\AA$.
\item
Col. 9: The observed H${\beta}$ fluxes in erg\,${\rm sec}^{-1}{\rm cm}^{-2}$,
corrected for absorption.
\item
Col. 10: Seeing corrected approximate angular size of the star-forming regions
r$_{HII}$, as measured in the B band by Vennik et al. (2000).
\item
Col 11: The morphological type.
\end{itemize}
Table 4 contains the corrected line ratios (normalised to H${\beta}=100$) of 
the emission lines detected in
the spectra of the galaxies listed in Table 3. The header of the table contains
the identification of the emission line while the line below gives the
corresponding restframe wavelength. The [SII]\,${\lambda}{\lambda}$6724,6731
and the [OII]\,${\lambda}{\lambda}$7320,7330  are not deblended, and therefore
the line ratios refer to their blend.

Table 5 lists the galaxies for which it was not possible to deblend 
H${\alpha}$ from [NII]\,$\lambda\lambda$\,6563,6584, and therefore the line 
ratios were not corrected for internal extinction. The line ratio of 
H${\alpha}$ corresponds to the blend with the [NII] lines. Finally, Table 6 
lists those galaxies for which the H${\beta}$ emission line was too noisy to 
allow an accurate determination of the internal extinction, and again only 
the uncorrected line ratios are listed. Here, the line ratios in
column 5 through 8 are given in units of H$\alpha$ = 100.
								
In 8 cases it was possible to disentangle the contribution of different HII
regions in the galaxy. Then the spectrum of each region was extracted
separately and the fluxes measured for each individual region. In Tables 3-6
only the line ratios corresponding to the brightest HII region are given, but
the H${\beta}$ fluxes are the total fluxes for the galaxy, summed over all the
HII regions. In Table 7 we list the 8 galaxies with detected HII regions
together with their observed H${\beta}$ fluxes. These fluxes are not corrected
for absorption.  The HII regions are given in order of their location along the
slit, from South to North.
\begin{table}[htb]
{\small
\begin{tabular}{ccccccc}
\multicolumn{5}{c}{{\bf Table 7 }}\\
\multicolumn{5}{c}{}\\
\hline\hline
& \\
Galaxy & \multicolumn{4}{c} {F(H${\beta})_{obs}$ [erg\,sec$^{-1}$\,cm$^{-2}$]}\\
 & HII1 & HII2 & HII3 & HII4  & \\
\hline
&\\

HS1232+3846 & 4.25e-16 & 2.33e-15 & 4.19e-16 & 3.62e-16 \\
HS1236+3821 & 5.01e-15 & 1.90e-16 \\
HS1244+3648 & 1.25e-15 & 1.39e-14 \\
HS1304+3529 & 2.57e-15 & 4.42e-14 & 1.45e-15 \\
HS1312+3508 & 2.12e-15 & 1.00e-13 \\
HS1340+3307 & 6.20e-16 & 9.11e-15 \\
HS1614+4709 & 5.92e-14 & 1.17e-15 & 5.81e-15 \\
HS1641+5053 & 1.37e-15 & 1.73e-14 \\
\hline
\end{tabular}
}
\end{table}

\section{Wolf-Rayet galaxies}

Some of our galaxies show the broad HeII ${\lambda}$4686 feature, which is a
direct signature from Wolf-Rayet (WR) stars. For low resolution spectroscopy 
(as in the case of our spectra) this
line is usually seen as a WR bump, a blend of HeII and other broad stellar
emission lines from CIII, NIII and NV, but also of narrow nebular emission
lines including HeII ${\lambda}$4686. The galaxies with detected WR features
are listed in Table 8 with the flag D (detection), while a few more candidates
for WR galaxies are also listed in Table 8, with the flag C (candidates). For
some of the galaxies with detected WR features we also list the
line ratios (corrected for internal extinction) relative to H$\beta$=100. The
first galaxy from the table, HS0915+5540, does not belong to the
spectrophotometric catalogue presented in this paper, but to the larger 
catalogue 
of emission-line galaxies from which the present catalogue was 
selected (Popescu et al. 1996). 

\begin{table}[htb]
{\small
\begin{tabular}{lll}
\multicolumn{3}{c}{{\bf Table 8 }}\\
\multicolumn{3}{c}{}\\
\hline\hline
& & \\
Galaxy & & Intensity$^a$\\
\hline
&&\\
HS0915+5540 & D & \\
HS1304+3529 & D & 5 \\
HS1312+3847 & C & \\
HS1312+3508 & D & 4\\
HS1424+3836 & D & 9\\
HS1442+4250 & C & \\
HS1507+3743 & D & 6\\
HS1657+5735 & D & 4\\
HS1728+5655 & C & \\
\hline
&&\\
\end{tabular}
}\\
$^a$ Dereddened line fluxes relative to I(H${\beta}$)=100.
\end{table}

In total we detected 6 new WR galaxies and 3 candidates for WR galaxies. This
represent 5.5$\%$ from our spectrophotometrical catalogue, or 9$\%$, if we
include the candidates, too. Such a  detection rate is somewhat lower than the
detection rate of 10$\%$ obtained by Masegosa et al. (1991) for a systematic search for
broad WR bump in the HII galaxies. However our 
spectral resolution is very low and it is therefore not optimised for this 
kind of search, and these results came only serendipitously.

The total number of known WR galaxies and extragalactic HII regions showing 
broad  HeII ${\lambda}$4686 emission is 139 (from the  recent compilation of 
Schaerer et al. (1999)). These objects are 
found among a large variety of morphological types, from BCDs and Im 
galaxies to massive spirals and luminous IRAS galaxies, and even in Seyfert 2 
and LINERSs (Osterbrock \& Cohen 1982, Kunth \& Contini 1998). Extending 
the sample of WR galaxies is very useful in constraining the evolution of 
massive stars and the parameters of the upper part of the IMF. Also, as 
discussed by Schaerer et al. (1999), WR galaxies represent useful templates 
of young starburst galaxies, which may be used to explain the properties of the distant
star-forming galaxies. 

Among the galaxies with detected WR features we found one SBN galaxy 
(HS1657+5735), a DANS galaxy (HS0915+5540), three DHIIH galaxies 
(HS1312+3508, \newline HS1424+3836,  HS1507+3743), and a interacting pair, 
IP (HS1304+3529). The candidates for WR features are SS, HIIH and DHIIH
galaxies. Thus it is probably reasonable to conclude that WR phenomenon is
spread over all morphological subtypes of BCDs. 

\section{Candidates for very low metallicities}

There has been a long debate on whether the BCDs are truly young dwarf
galaxies undergoing their first burst of star formation or whether the present
burst occurs within an older galaxy. The low abundances found in these galaxies
make them good candidates for the least chemically evolved galaxies. But
the detection of an extended faint stellar underlying component in the
majority of BCDs (Loose \& Thuan 1986, Kunth et
al. 1988, Telles et al. 1997, Vennik et al. 1999) 
supports the idea that they are not truly 
primordial galaxies, but older LSB dwarf galaxies undergoing transient 
periods of star formation. A clear demonstration of this two - component 
structure was recently presented by Schulte-Ladbeck et al. (1998, 1999) for 
the very nearby BCD VII Zw 403 (UGC 6456), based on HST photometry of its individual 
stars down to the red giant branch stars. Even for I Zw 18, the lowest 
metallicity known galaxy, Garnett et al. (1997) found C/O to be much larger 
than the mean value of other metal-deficient galaxies. This was interpreted 
as a proof that I Zw 18 has experienced carbon-enriching episodes of star 
formation in the past, and is therefore not a young galaxy. Recently 
Aloisi et al. (1999) used synthetic colour-magnitude diagrams to investigate 
deep HST data for I Zw 18, and they found again that the present burst is not 
the first one to occur in this galaxy. 

However, Izotov \& Thuan (1999) argued that the extreme low metallicity BCD
SBS 0335-052 is a good example for a young galaxy; the HST V and I imaging of
this galaxy (Thuan et al. 1997) showed extremely blue colours, 
not only in the region of current star formation but also in the extended low 
surface brightness underlying component. Thuan \& Izotov (1997) also 
argued  that the large HI cloud associated with this BCG (Pustilnik et
al. 1999) is made of pristine gas, unpolluted by metals.  Based on
abundance measurements, Izotov \& Thuan (1999) suggested that in fact all
galaxies with 12+log(O/H)$\leq7.6$  are young, with ages not exceeding 40 Myr,
while those with $7.6 < 12+{\rm log}({\rm O}/{\rm H}) < 8.2$ are not older 
than 1-2 Gyr. Furthermore, Lynds et al. (1998) discussed the possibility 
that VII Zw 403 (UGC 6456) is not young (as also found by Schulte-Ladbeck 
et al. 1998, 1999), but could still be considerably younger 
than a Hubble time.

While the controversy on the age of the BCDs reflects our current knowledge of
galaxy formation and evolution (Schulte-Ladbeck et al. 1999), more work on both
statistical samples of very low-metallicity BCDs as well as HST imaging of
nearby BCDs could give clues relevant to this debate. In this section we give a
list of candidates for very-low metallicity BCDs, which were found in the
spectroscopical analyses of our sample. Nevertheless, high-resolution
spectroscopy is needed to confirm the metallicity that we assign for each 
object.

\begin{figure}[htp]
\includegraphics[scale=0.4]{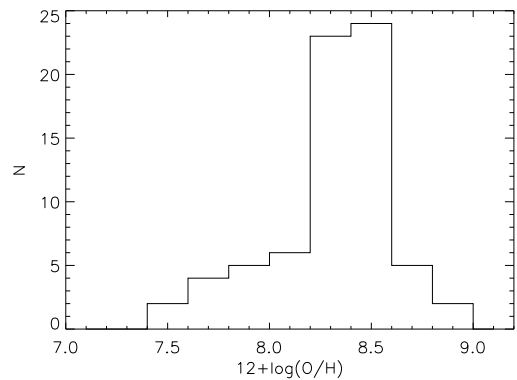}
\caption{The distribution of the oxygen abundances 12+log(O/H).} 
\end{figure} 

We derived the metallicities for our sample using both the models of Dopita \&
Evans (1986) as well as a five level atom program and the ionisation correction
method of Mathis \& Rosa (1991), the latter for 
the few objects for which the [OIII]\,${\lambda}$4363 line was detected in the
spectrum. The distribution of 12+log(O/H) (Fig. 1) has a maximum around 8.5,
with a long tail towards low and very low metallicities. All the galaxies with
metallicities less than 8.0 were considered candidates for very low 
metallicity objects and are listed in Table 9. The value of the metallicity 
we assign should be taken with caution, and only as a preliminary result. 
As a simple consistency check, we used the data from Table 9 to plot the
metallicity - luminosity relation of star forming dwarf
galaxies, as given by Skillman et al. (1989). Since the dwarf galaxies used by
Skilmann et al. are all dwarf irregular galaxies, we further added some 
BCD's from Thuan et al. (1997). Our data distribute in the same region as the 
Thuan et al. BCDs, with a similar amount of scatter and are - within the 
errors - in good agreement to the relation proposed by Skillman et
al. (1989)(see Fig. 2). We conclude that the errors in
our preliminary 12+log(O/H) determinations are small enough to
indicate the existence in our sample of some interesting low-abundance
BCDs. However, we can not exclude a small offset in our O/H scale.

\setcounter{table}{8}
\begin{table}[htb]
\caption{Candidates for very low metallicity galaxies}
{\small
\begin{tabular}{lll}
\hline\hline
&  \\
Galaxy & 12+log(O/H) & M$_{B}$\\
\hline
&\\
HS1222+3741 &  7.64& -17.67\\
HS1236+3937 &  7.47& -15.67\\
HS1304+3529 &  7.66& -17.38\\
HS1318+3239 &  7.81& -17.22\\
HS1319+3224 &  7.59& -15.3\\
HS1330+3651 &  7.66& -17.15\\
HS1347+3811 &  7.94& -15.30\\
HS1424+3836 &  7.97& -16.06\\
HS1440+3120 &  7.91& -17.4\\
HS1442+4250 &  7.89& -14.73\\
HS1507+3743 &  7.74& -17.58\\
\hline
&\\
\end{tabular}
}\\
\end{table}

\begin{figure}[htp]
\includegraphics[scale=0.4]{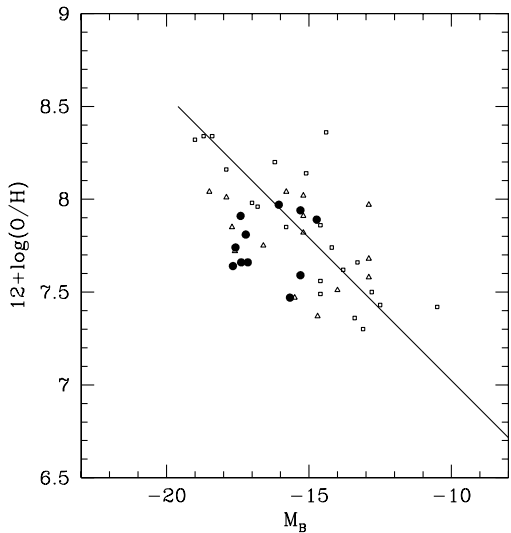}
\caption{The oxygen abundances (12+log(O/H)) - absolute magnitude
relation as derived by Skillman et al. (1989) for the local dwarf
irregular galaxies (open squares). Some selected BCDs from Thuan et
al. (1997) are also plotted (open triangles) together with our 
low-abundance candidates (filled dots).} 
\end{figure} 

\section{The properties of the sample}

In this section we present some statistics on the relevant spectroscopic
parameters of the sample as well as of subsamples of different morphological
subtypes.

Our sample was selected in certain fields which contain huge
underdensities. One concern is that its galaxy content is thus
different from the galaxy content of samples selected from the general field 
environment. In passing we should note that Popescu et al. (1997) found that 
most of our sample galaxies belong to the normal field features of the large 
scale structure. Nevertheless, we make an extra check, by comparing the 
frequency distribution of the Salzer et al. (1989b) types (see Tables 3, 5, and
6) with those from the University of Michigan (UM) survey, which is a general 
field survey. The comparison is given in Table 10 and indicates little significant
differences in the type distribution of the two samples. If at all, we
have some more DHIIH galaxies and fewer SBN objects in our sample. This is to
be expected due to the selection criteria we adopted, namely bright objects
were excluded from the survey (Popescu et al. 1996). The motivation
was that bright objects were already included in other catalogues and we were
mainly interested in dwarf galaxies. Since the frequency distribution of
different morphological types is very close to that of the UM survey, we also 
expect a correlation between the Salzer et al. (1989) morphological types 
of our objects and their absolute blue magnitude. We verified that this 
correlation exists, but with a reasonable scatter in each sub-class. This 
scatter (peak-to peak) can be as large as 4 mag in most of the type bins.
\setcounter{table}{9}
\begin{table}[htb]
\caption{Frequency distribution of the ELG
types in the University of Michigan survey (UM) and in our sample.}
{\small
\begin{tabular}{lrr}
\hline\hline
&  \\
Type & this paper & UM\\
     & [\%] & [\%] \\
\hline
&\\
SS    &  4.8 $\pm$ 2.4 &  9.9 \\ 
DHIIH & 39.3 $\pm$ 6.8 & 29.7 \\
HIIH  & 26.2 $\pm$ 5.6 & 24.8 \\
DANS  & 14.3 $\pm$ 4.1 & 12.4 \\
SBN   &  8.3 $\pm$ 3.1 & 13.2 \\
IP    &  7.1 $\pm$ 2.9 &  9.9 \\
\hline
&\\
\end{tabular}
}\\
\end{table}
For a detailed discussion of the sample
properties as a function of the galaxy density see Popescu et al. (1999).

\begin{figure*}[htp]
\plotfiddle{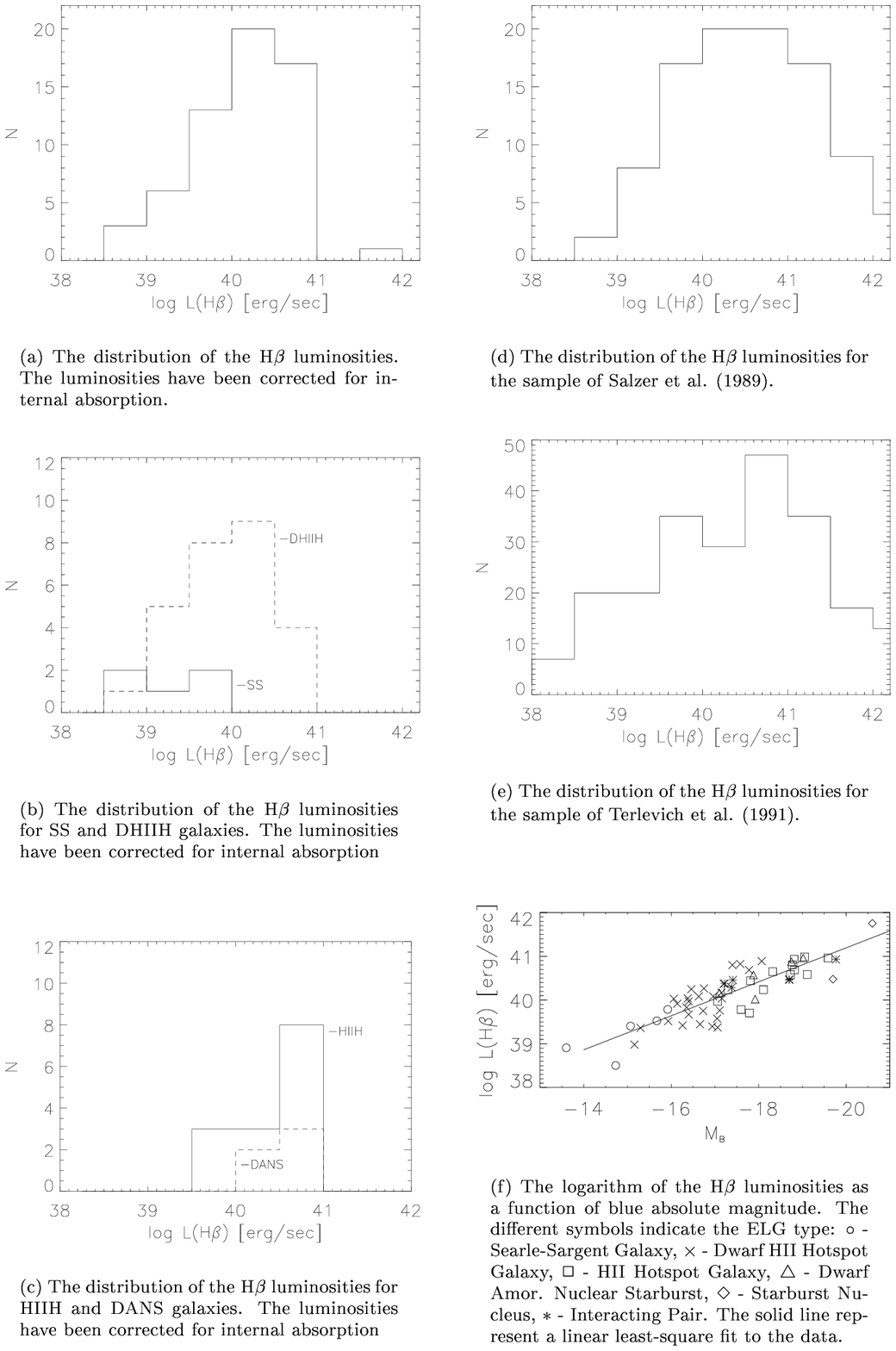}{9.0in}{0.}{100.}{100.}{-350.}{-80.}
\caption[]{} 
\end{figure*} 

The histogram of the H${\beta}$ luminosities of the whole sample is given in 
Fig. 3a. The luminosities were
calculated for a Hubble constant ${\rm H0} =75$\,km/s/Mpc and they were
corrected for internal extinction. The distribution is asymmetric, with a sharp
cut-off towards the brighter end. This is due to the selection criteria we
adopted, as mentioned above (see also Popescu et
al. 1996). The 
incompleteness at the bright end is also obvious when compared
with the H${\beta}$ distribution of other similar samples from the 
literature. For example, in Fig. 3d we show, for comparison, the H${\beta}$ 
luminosity distribution of a sample of 99 ELGs (Salzer et al. 1989b) from the 
UM objective-prism survey and in Fig. 3e the same 
distribution from the ELG sample of Terlevich et al. (1991). In comparison with
our distribution, the latter ones are more symmetrical and span over more
orders of magnitude. They include at the bright end also Seyfert galaxies, 
which were not included in our study. The H${\beta}$ luminosity distributions 
of different morphological subtypes are shown in Fig. 3b (for SS and DHIIH
galaxies) and in Fig. 3c (DANS and HIIH), respectively. The differences between
them show the known trend of increasing H${\beta}$ luminosities from
the SS to DANS class, though a significant overlap exist, too.

The expected correlation between the H${\beta}$ luminosities and the blue
magnitudes is shown in Fig. 3f, where different morphological subtypes are
plotted with different symbols. A linear least square fit to the
log(L(H${\beta}$))=f(M$_B$) data  
(plotted with solid line) gives a slope of -0.388 and a correlation 
coefficient of 0.85. This is close to the
the slope (-0.391) of a similar correlation found by Salzer et al. (1989b) 
for the ELGs of the Michigan Survey. The slope of -0.391 was derived when the
Seyfert and SS galaxies were excluded from the correlation. This is consistent
with our set of data, which does not contain Seyfert galaxies. Also the SS
galaxies of our sample seem to follow in a better way the correlation, while
those found by Salzer are lying mainly in the upper part of the correlation. As
remarked by Salzer, this slope suggests that, within the uncertainties, the
H${\beta}$ luminosity scales directly with the blue luminosity: 
L(H${\beta})\propto {\rm L}_B$. This would indicate that the recent
star-formation (in the last $\sim10$ Myr) and the integrated star formation 
are related. On the other hand the scatter from the correlation suggests
variations in the global equivalent widths, W(H$\beta$), which means different
stages of activity. If the total blue magnitude is a good measure of the
average past star formation, then the scatter indicates an intrinsic variation
in the ratio between the present star formation rate and the average past.

\begin{figure}[htp]
\includegraphics[scale=0.4]{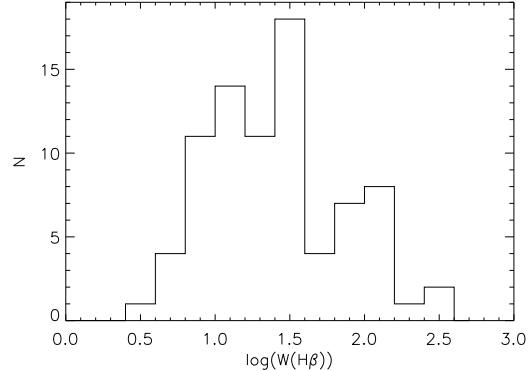}
\caption{The distribution of the H${\beta}$ equivalent widths 
(in units of $\AA$. } 
\end{figure} 

The distribution of the H${\beta}$ equivalent widths is also given in Fig. 4.

\section*{Appendix}

During our survey for emission line galaxies we also identified a few quasars
and AGNs. Here we give some spectroscopic information about a peculiar QSO
found in our survey, namely HS1643+5313. This object looked peculiar 
also in the objective-prism scans, in the sense that it displayed a double
peaked feature, just at the green head of the objective prism spectrum. A
first possibility
would have been that the strongest feature were  the [OIII]${\lambda}$5007 
line of a narrow-emission line galaxy while the second feature a cosmic or 
only noise. No
other combination of emission-lines at any redshift and at this dispersion
were known to fit the above mentioned spectrum. The spectrum was therefore
chosen
mainly with the hope of being still a narrow-emission line galaxy. The slit
spectrum (Fig. 5) displays the same combination of
a double emission-line feature. After analysing in detail the spectrum as well
as its direct image we reached the conclusion that the object is a QSO at
z=0.785, the emission-features being thus only one line, namely the 
MgII\,${\lambda}{\lambda}$2798 blend,
but with an absorption dip
inside. There is also some faint detection of FeII at 5262\,\AA, but the
S/N ratio of the spectrum is too poor for a clear detection of further fainter 
lines.
 A high
resolution spectrum of the near infrared region would clarify whether 
the absorption is internal to the QSO or arises from a foreground cloud or
galaxy.

\begin{figure}[htp]
\includegraphics[scale=0.4]{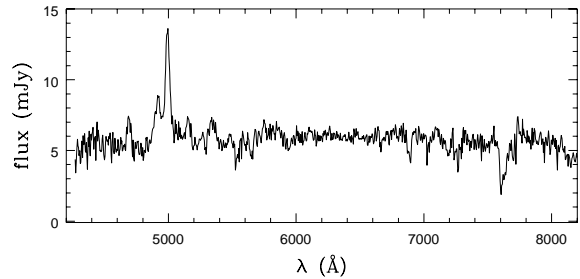}
\caption{Slit spectrum of the quasar HS1643+5313, z=0.785.} 
\end{figure} 

\begin{acknowledgements}
U.H. acknowledges the support of the SFB 375.

This research has made use of the NASA/IPAC Extragalactic Database (NED) which
is operated by the Jet Propulsion Laboratory, California Institute of
Technology, under contract with the National Aeronautics and Space 
Administration.

\end{acknowledgements}

\end{document}